\documentclass[sigconf]{acmart}

\copyrightyear{2020}
\acmYear{2020}
\setcopyright{iw3c2w3}
\acmConference[WWW '20]{Proceedings of The Web Conference 2020}{April 20--24, 2020}{Taipei, Taiwan}
\acmBooktitle{Proceedings of The Web Conference 2020 (WWW '20), April 20--24, 2020, Taipei, Taiwan}
\acmPrice{}
\acmDOI{10.1145/3366423.3380285}
\acmISBN{978-1-4503-7023-3/20/04}

\usepackage[utf8]{inputenc}
\usepackage[export]{adjustbox}
\newcommand{\balpha}{\boldsymbol{\alpha} }
\newcommand{\bbeta}{\boldsymbol{\beta} }
\newcommand{\bgamma}{\boldsymbol{\gamma} }
\newcommand{\bX}{\boldsymbol{X} }
\newcommand{\bH}{\boldsymbol{H} }

\newcommand{\bT}{\boldsymbol{T} }
\newcommand{\bQ}{\boldsymbol{Q} }
\newcommand{\bK}{\boldsymbol{K} }
\newcommand{\bV}{\boldsymbol{V} }

\usepackage{multirow}
\usepackage{tikzsymbols}
\usepackage{enumitem}

\begin{document}

\title{Déjà vu: A Contextualized Temporal Attention Mechanism for Sequential Recommendation}
% \title{Time is What You Cannot Miss}


\author{Jibang Wu}
\affiliation{%
  \institution{University of Virginia}
%   \streetaddress{}
  \city{Charlottesville}
  \state{VA}
  \country{USA}}
\email{jw7jb@virginia.edu}

\author{Renqin Cai}
\affiliation{%
  \institution{University of Virginia}
%   \streetaddress{}
  \city{Charlottesville}
  \state{VA}
  \country{USA}}
\email{rc7ne@virginia.edu}

\author{Hongning Wang}
\affiliation{%
  \institution{University of Virginia}
%   \streetaddress{}
  \city{Charlottesville}
  \state{VA}
  \country{USA}}
\email{hw5x@virginia.edu}
% \renewcommand{\shortauthors}{Jibang Wu, et al.}

\begin{abstract}
Predicting users' preferences based on their sequential behaviors in history is challenging and crucial for modern recommender systems. Most existing sequential recommendation algorithms focus on transitional structure among the sequential actions, but largely ignore the temporal and context information, when modeling the influence of a historical event to current prediction. 

In this paper, we argue that the influence from the past events on a user's current action should vary over the course of time and under different context. Thus, we propose a Contextualized Temporal Attention Mechanism that learns to weigh historical actions' influence on not only what action it is, but also when and how the action took place. More specifically, to dynamically calibrate the relative input dependence from the self-attention mechanism, we deploy multiple parameterized kernel functions to learn various temporal dynamics, and then use the context information to determine which of these reweighing kernels to follow for each input. In empirical evaluations on two large public recommendation datasets, our model consistently outperformed an extensive set of state-of-the-art sequential recommendation methods.
\end{abstract}




%   \includegraphics[width=\textwidth]{img/motivating_example.jpg}

\maketitle
% \section*{Outline}

\section{Introduction}
% \textbf{[What is sequential recommendation and why is it important?]}
The quality of recommendation results is one of the most critical factors to the success of online service platforms, with the growth objectives including user satisfaction, click- or view-through rate in production. Designed to propose a set of relevant items to its users, a recommender system faces dynamically evolving user interests over the course of time and under various context. For instance, it is vital to distinguish when the history happened (e.g. a month ago or in the last few minutes) as well as to evaluate the context information (e.g. under a casual browsing or some serious examining setting), especially on how serious the user is about the click, and how related his/her preference is to this particular event.
% context. 

\begin{figure}[t]
  \centering
  \includegraphics[width=\linewidth, left]{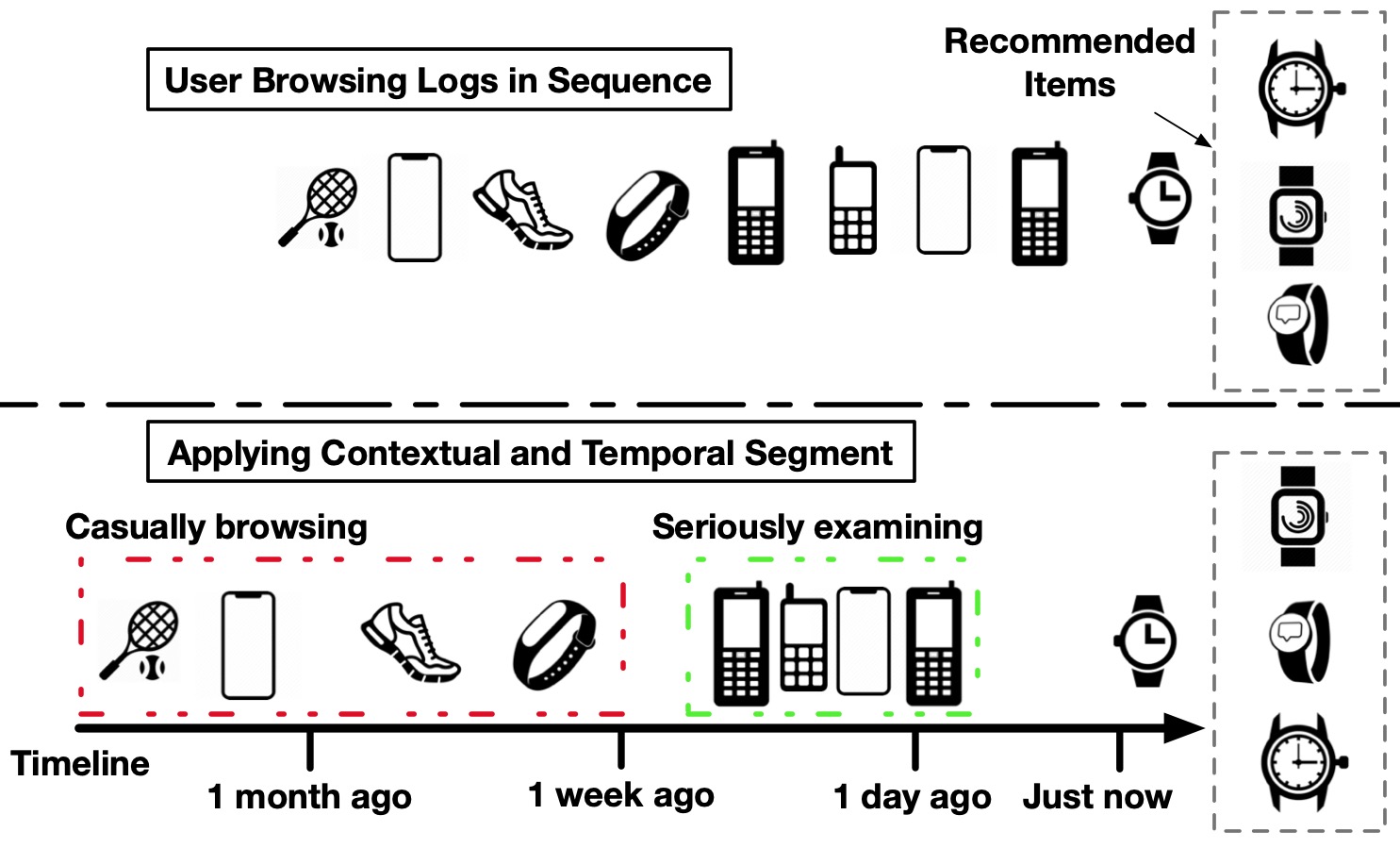}
  \caption{Sequential recommendation in online shopping scenario (up) from the traditional view, (down) from our view with temporal and contextual segmentations.  } %The timeline is presented in non-linear scale to showcase 
  \label{fig:motiv_exp}
  \vspace{-4mm}
\end{figure}

Concerning such sequential dependence within user preferences, the task of sequential recommendation is set to predict the ongoing relevant items based on a sequence of the user's historical actions. 
Such setting has been widely studied \cite{hidasi2015session, quadrana2017personalizing, you2019hierarchical, cui2017hierarchical, kang2018self, Tang:2018:PTS:3159652.3159656, chen2018sequential, cai2018modeling} and practiced in popular industry recommender systems such as YouTube \cite{tang2019towards, beutel2018latent} and Taobao \cite{sun2019bert4rec}. Take the online shopping scenario illustrated in Figure \ref{fig:motiv_exp} for example: the system is given a series of user behavior records and needs to recommend the next set of items for the user to examine. We should note in this setting we do not have assumptions about how the historical actions are generated: solely from interaction between the user and the recommender system, or a mix of users' querying and browsing activities. But we do assume the actions are \emph{not} independent from each other. This better reflects the situation where only offline data and partial records of user behaviours are accessible by a recommender system. 
%then displays a collection of recommended items.

 % What is the challenge facing in sequential recommendation setting
 One major challenge to the sequential recommendation task is that the influence patterns from different segments of history reflect user interests in different ways, as is exemplified in Figure \ref{fig:motiv_exp}: 
 \begin{itemize}[leftmargin=*]
     \item By \textit{temporal segment}: The distant history indicates that the user is interested in shopping sports related products. Now that he or she is looking for a watch, the system could have recommended some sports watches instead of generic ones. Essentially, the distant and prolonged user history could carry sparse yet crucial information of user preferences in general, while the more recent interactions should more closely represent the user intention in near future.
     \item By \textit{contextual segment}: Since the user closely examined several smartphone options (much shorter time intervals in between than the average), these interaction events could be emphasized for estimating current user preference such that smartwatches might be preferred over traditional watches. In general, some periods of user browsing log could appear to be heterogeneous, packed with exploration insights, while at a certain point, the user would concentrate on a small subset of homogeneous items, in a repetitive or exploitative way. 
 \end{itemize}
Hence, the designs to capture and connect these different signals from each part of history have driven the progress of recent development of sequential recommendation algorithm. 
 
 % What is the existing solution to this challenge?
 Traditionally, the notion of session is introduced in modeling sequential user behaviors, in a way to segment sequence of actions by active and idle engagement. It is shown in \cite{hidasi2015session, quadrana2017personalizing} that the pattern of user preference transition usually differs after a session gap, which is commonly defined as a minimal of 30 minutes' inactivity. Prior work has demonstrated many effective hierarchical neural network structures, such as hierarchical recurrent neural networks (RNN) \cite{quadrana2017personalizing, cui2017hierarchical, you2019hierarchical} to jointly model the transition patterns in- and cross- sessions. However, the strength stemmed from its session assumption could also be the bottleneck of session based recommendation, that is, the user preference does not necessarily transit in strict accordance with the manually defined session boundaries. 
 
 \begin{figure}[H]
  \centering
  \includegraphics[width=0.48\linewidth]{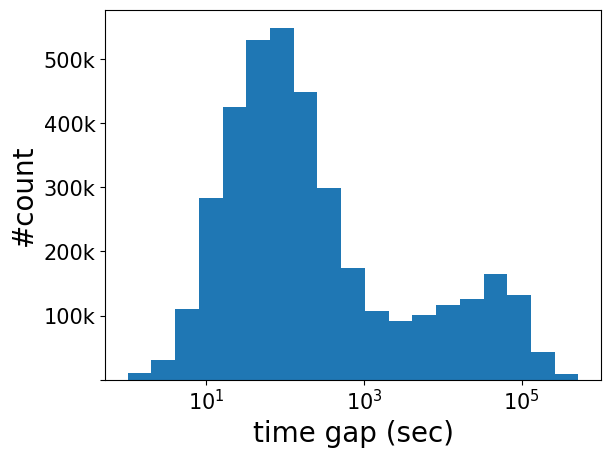}
  \includegraphics[width=0.48\linewidth]{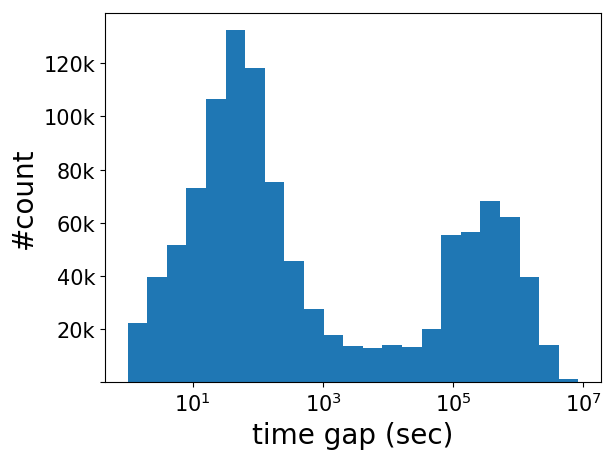}
  \caption{Histogram of time intervals between successive events on two datasets, UserBehavior (left), XING (right).}
  \label{fig:stat_exp}
  \vspace{-4mm}
\end{figure}
  %  \textbf{[Why would action time be useful information to model in sequential recommendation, especially joint with contextual information?]}
 In this paper, we argue that the transition patterns could widely differ due to the subtle variance within the temporal proximity between neighboring events, associated with its changing context. 
 %% show statistical result if possible   %figure of time distribution ??? %figure for context information??
 Specifically, the time and context of each historical event would support fine-grained interpretation and high-fidelity replay of the sequential behavior history for a more accurate portrait of current user preference. This claim is further supported by the initial statistic results obtained from two large datasets later used in our evaluation: the mixed Gaussian shape appearing in the time interval distribution in Figure \ref{fig:stat_exp} indicates that a binary interpretation of time gap as in- or cross- session is not accurate enough. Therefore, our proposed model adaptively weights the historical influences in regard to the user's drifting impressions from previous interactions over time and under its inferred contextual condition. 
 %As the online shopping example in Figure \ref{fig:motiv_exp}, where the user's actions are spread across the timeline, the user decides the relevant item at current moment in regard to his or her drifting impressions from previous interactions over time and under certain contextual condition. 
 %as oppose to the common choice of RNN based deep learning approaches in sequential behaviour modelling\cite{hidasi2015session,quadrana2017personalizing}

 %%% why do we base our method on attention? what the connection between our definition of history segment  and attention mechanism?
 Traditional RNN-based approaches leave little room for one to dynamically adjust the historical influences at current state. One earlier work, known as Time-LSTM \cite{zhu2017next}, proposed several variants of time gate structure to model the variable time intervals as part of the recurrent state transition. But this assumes that the temporal influence only takes effect for once during transition and is fixed regardless of context or future events. Thereby, in order to model the influence evolving temporally and contextually, we appeal to the attention based sequence models, which emphasize dependencies directly on each sequential input rather than relying on the recurrent state transition in RNN models. 
 
 In sequential recommendation, a line of work  \cite{kang2018self, tang2019towards, sun2019bert4rec} has borrowed the state-of-art self-attention network structure from nature language modeling \cite{devlin2018bert, vaswani2017attention}. \citet{tang2019towards} show that its attention component can enhance the model capacity in determining dependence over an extensively long sequence of history. Nevertheless, Kang and McAuley \cite{kang2018self} report that the action order in long sequence of user interaction history is lacking in boosting the empirical evaluation performance on several recommendation datasets, even though the position embedding technique is proposed for self-attention mechanism in its original paper \cite{vaswani2017attention}. %%% more investigation on position embedding
 In other words, there is no explicit ordering of input or segment of history modeled by self-attention mechanism. Therefore, this presents us the opportunity to model temporal and context information as a more informative and flexible order representation to complement existing attention mechanism, bridging the insights from the both sides of work in sequential recommendation. But the challenge also comes along as incorporating these information could  contribute more noise than signal unless properly structured by the model.

 % outline of the model
 We propose the \textbf{C}ontextualized \textbf{T}emporal \textbf{A}ttention Mechanism (\textbf{CTA}), an attention based sequential neural architecture that draws dependencies among the historical interactions not only through event correlation but also jointly on temporal and context information for sequential behavior modeling. In this mechanism, we weigh the historical influence for each historical action at current prediction following the three design questions:
 \begin{enumerate}
    \item \textit{What is the action?} The dependency is initially based on the action correlation through the self-attention mechanism, i.e., how such an action is co-related to the current state in the sequence.
    \item \textit{When did it happen?} The influence is also weighed by its temporal proximity to the predicted action, since the temporal dynamics should also play an important role in determining the strength of its connection to presence.
    \item \textit{How did it happen?} The temporal weighing factor is realized as a mixture of the output each from a distinct parameterized kernel function that maps the input of time gaps onto a specific context of temporal dynamics. And the proportion of such mixture is determined by the contextual factors, inferred from the surrounding actions. In this way, the influence of a historical action would follow different temporal dynamics under different contextual conditions.
 \end{enumerate}

We apply the model on both XING\footnote{https://github.com/recsyschallenge/2016/blob/master/TrainingDataset.md} and UserBehavior\footnote{https://tianchi.aliyun.com/dataset/dataDetail?dataId=649} dataset each with millions of interaction events including user, item and timestamp records. The empirical results on both dataset show that our model improves recommendation performance, compared with a selection of state-of-the-art approaches. We also conducted extensive ablation studies as well as visualizations to analyze our model design to understand its advantages and limitations.

% outline of contributions in the paper
% \textbf{Contributions:} We first present our perspective of the sequential recommendation problem with the extra information of interaction time and the event context. We then propose the Contextualized Temporal Attention Mechanism, an attention based sequential neural architecture that draws dependencies among the historical interactions not only through event correlation but also jointly on temporal and context information for sequential behaviour modeling task. Essentially, the attention on each historical event is determined by its content based correlation with current state from the self-attention mechanism, as well as its temporal order through a mixture of signals from multiple temporal kernels assorted by the context. We apply the model on both XING and Taobao dataset each with millions of interaction events including user, item and timestamp records. The empirical results on both dataset show that our model improves recommendation performance, compared with a selection of existing approaches.

\section{related work}
Several lines of existing research are closely related to ours in this paper, and their insights largely inspired our model design. In this section, we briefly introduce some key work to provide the context of our work.

\subsection{Sequential Recommendation}

%\textbf{Session-based recommendation.}
%Deep neural networks have gained tremendous success in the fields of Computer Vision (CV)  and Natural Language Processing (NLP)\cite{lecun2015deep, manning2015computational}. In recommendation algorithm research, there is also growing interest of using deep neural networks to model complex interactions between user and items, some of which\cite{yu2015multiscale, Sedhain:2015:AAM:2740908.2742726} has surpassed classic matrix factorization-based methods \cite{Koren:2009:MFT:1608565.1608614, Linden:2003:ARI:642462.642471}. Recent work \cite{He:2017:NCF:3038912.3052569} also proved that traditional Collaborative Filtering methods can be effectively generalized by a deep neural network. 
For the problem of sequential recommendation, the scope was initially confined to the time-based sessions. Recurrent Neural Network (RNN) and its variants, including Long Short-Term Memory (LSTM) \cite{hochreiter1997long} and Gated Recurrent Units (GRU) \cite{chung2014empirical}, have become a common choice for session-based recommendations \cite{Zhang:2014:SCP:2893873.2894086, hidasi2015session}. Other methods based on Convolutional Neural Networks (CNN) \cite{Tang:2018:PTS:3159652.3159656}, Memory Network \cite{chen2018sequential} and Attention Models \cite{li2017neural} have also been explored. The hierarchical structure generalized from RNN, Attention or CNN based models \cite{cui2017hierarchical, quadrana2017personalizing, you2019hierarchical} is used to model transitions inter- and intra-sessions. The recent work \cite{you2019hierarchical} by You et al. showed that using Temporal Convolutional Network to encode and decode session-level information and GRU for user-level transition is the most effective hierarchical structure. Nevertheless, as many studies borrow sequence models from natural language modeling task 
%to capture similar sequential patterns in sequential recommendation
directly, their model performance is usually limited by the relatively small size and sparse pattern of user behaviors, compared to the nature language datasets. 
%It is found that these sequential model performs better in certain subsequence such as session\cite{hidasi2015session}, where the actions more closely related. 
%Therefore, it is crucial to develop sequential model that captures the recommendation specific pattern to better solve this problem.

%\textbf{Attention-based sequential recommendation.}
The attention mechanism was first coined by \citet{bahdanau2014neural}. The original structure is constructed on the hidden states generated from RNN in order to better capture the long-term dependence and align the output for decoder in RNN. 
%Attention mechanism is shown to be effective in various tasks such as machine translation \cite{luong2015effective}, image captioning \cite{xu2015show} and etc \cite{chorowski2015attention, yang2016hierarchical}. 
The Transformer model \cite{vaswani2017attention} and several follow-up work \cite{devlin2018bert, dai2019transformer, radford2018language} showed that for many NLP tasks, the sequence-to-sequence network structure based on attention alone, a.k.a. self-attention mechanism, is able to outperform existing RNN structures in both accuracy and computation complexity in long sequences. Motivated by this unique advantage of self-attention, several studies introduced this mechanism to sequential recommendation. 
%Many existing methods developed for sequential recommendations have the limitation especially when dealing with long user sequences found in many production recommender systems. 
SASRec \cite{kang2018self}, based on self-attention mechanism, demonstrated promising results in modeling longer user sequences without the session assumption. Another work known as Multi-temporal range Mixture Model (M3) \cite{tang2019towards} manages to hybrid the attention and RNN models to capture the long-range dependent user sequences. The most recent work, BERT4Rec \cite{sun2019bert4rec}, adopts the bidirectional training objective via Cloze task and further improves its performance over SASRec. %In particular, the results from SASRec show that the position embedding only improves accuracy on sparse dataset but hurts on dense ones, where each behavior occurs in shorter time interval. This also motivates us directly use temporal information in attention score rather than only the abstract positional information.

%The self-attention mechanism takes the input of the key, query, value variables $K, Q, V$ and use the scaled dot product to calculate the weighted sum as the output on key, query and value tuples. 

%The self-attention mechanism is used for different stages: As an input encoder, $K, V, Q$ are produced from input vectors; As an output decoder, $K, V, Q$ are produced from output vectors; As an encoder to decoder (seq2seq) model, $K, V$ are produced from input vectors and $Q$ is produced from output vectors.

%The design of position embedding is also introduced in Transformer to inject the information of input sequence order, since the self-attention mechanism contains no recurrence and no convolution. 

\subsection{Temporal Recommendation}
%\textbf{Temporal information in recommendation.}
Temporal recommendation specifically studies the temporal evolution of user preferences and items; and methods using matrix factorization have shown strong performance. TimeSVD++ \cite{koren2009collaborative} achieved strong results by splitting time into several bins of segments and modeling users and items separately in each. Bayesian Probabilistic Tensor Factorization (BPTF) \cite{xiong2010temporal} is proposed to include time as a special constraint on the time dimension for the tensor factorization problem. And many of these solutions \cite{xiong2010temporal, song2016multi, li2014modeling} in temporal recommendation share the insight to model separately the long-term static and short-term dynamic user preference. Nevertheless, none of the models are developed specifically for sequential recommendation.

%\textbf{Learning to model temporal Information in Sequence.}
There have been various efforts to utilize temporal information in existing deep recommendation models. \citet{li2017time} proposed methods to learn time-dependent representation as input to RNN by contextualizing event embedding with time mask or event-time joint embeddings.  \citet{zhu2017next} proposed several variants of time gate structure to model the variable time interval as part of the recurrent state transition. But the empirical results of both model show limited improvement compared to their LSTM baseline without using temporal information.%, and one reason might be that these structures only model the time interval from the last event, limiting the amount of the temporal information the model can utilize.
% \begin{align*}
% x_{t} \gets x_{t} \odot m_{d} \quad & 
% \textup{Time mask: }  m_{d} =\sigma\left(c^{d} W_{d}+b_{d}\right), \quad c^{d} =\phi\left(\log \left(d_{t}\right) ; \theta\right) \\
% x_{t}  \gets \frac{1}{2} (x_{t}+g_{d}) \quad &
% \textup{Time embedding: } g_{d} =\sigma\left(d_{t} W_{d}+b_{d}\right) E^{s} 
% \end{align*}
%In Spoken language understanding (SLU), Su et al. \cite{su2018time} has explored the time decay attention based on Bi-LSTM when aggregating the multi-turn information in dialogues. The insight is that the most recent utterances are more important than the least recent ones. Its experiment shows significantly improved contextual understanding performance on the state-of-the-art model.

%\citet{chen2018neural} proposed the model, known as the neural ordinary differential equations, also envisioning a continuous modeling of hidden state. Their model learns the temporal variation of hidden state as solving multiple ordinary differential equations. While it is shown to be robust in learning latent function of time-series data on simulated datasets, we still have limited understanding in its effectiveness and scalability for real-world data.

Meanwhile, the time series analysis is a well established research area with broad application in real world problems \cite{das1994time, scharf1991statistical}. %Traditional statistical methods are still the more popular choices than deep neural networks, especially for noisy and lengthy time series data. 
Hawkes process \cite{hawkes1971spectra, laub2015hawkes} is one of the powerful tools for modeling and predicting temporal events. %with applications from earthquake modeling to financial analysis. 
It models the intensity of events to occur at moment $t$ conditioned on the observation of historical events.
%$$ \lambda\left(t | \mathcal{H}_{t}\right)=\lambda_{0}(t)+\sum_{i : t>T_{i}} \phi\left(t-T_{i}\right) $$
Some recent work \cite{vassoy2019time, xiao2017modeling, du2016recurrent} attempt to use RNN to model the intensity function of point process model and predict the time of next action. As a typical example, the Neural Hawkes Process \cite{mei2017neural} constructs a neurally self-modulating multivariate point process in LSTM, such that the values of LSTM cells decay exponentially until being updated when a new event occurs. Their model is designed to have better expressivity for complex temporal patterns and achieves better performance compared to the vanilla Hawkes process. The Long- and Short- Term Hawkes Process model \cite{cai2018modeling} demonstrates a combination of Hawkes Process model for different segments of user history can improve the performance in predicting the type and time of the next action in sequential online interactive behavior modeling. However, most of these Hawkes process based algorithms model each typed event as a separate stochastic process and therefore cannot scale as the space of event type grows. 

% \textit{Contextual Recommendation.}

% \textit{Session Recommendation.}

% temporal segment
% contextual segment
\section{method}
In this section, we discuss the details of our proposed \textbf{C}ontextualized \textbf{T}emporal \textbf{A}ttention Mechanism (\textbf{CTA}) for sequential recommendation. We will first provide a high-level overview of the proposed model, and then zoom into each of its components for temporal and context modeling.  

\subsection{Problem Setup \& Model Overview}
We consider the sequential recommendation problem with temporal information. Denote the item space as $\mathcal{V}$ of size $N$, and the user space as $\mathcal{U}$ of size $U$. The model is given a set of user behavior sequences $\mathcal{C} =\{ \mathcal{S}^1,\mathcal{S}^2, \dots, \mathcal{S}^U\}$ as input. Each  $\mathcal{S}^u = \{ (t_1^u, s_1^u) , (t_2^u, s_2^u) , \dots \}$ is a sequence of time-item tuples, where $t_j^u$ is the timestamp when item $ s_i^u$ is accessed by user $u$, and the action sequence is chronologically ordered, i.e., $t^u_i \leq t^u_{i+1} $. The interacted item is represented as a one-hot vector $ s\in \mathbb{R}^{1 \times N}$ and the timestamp is a real valued scalar $t\in \mathbb{R}^+$. The recommendation task is to select a list of items $  V \subseteq  \mathcal{V}$ for each user $u$ at a given time $t$ with respect to $\mathcal{S}^u$, such that $V$ best matches user $u$'s interest at the moment. 

\begin{figure}[t]
  \centering
  \includegraphics[width=\linewidth, left]{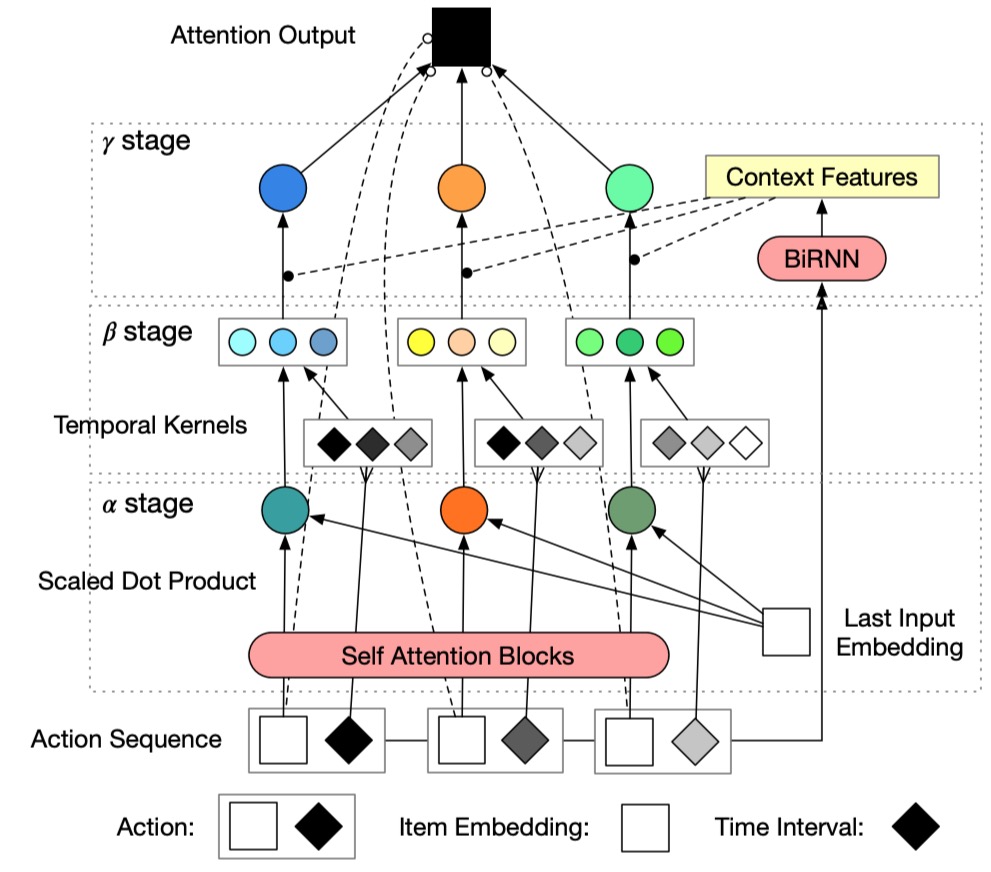}
  \caption{The architecture of our proposed Contextualized Temporal Attention Mechanism. Three stages are proposed to capture the content information at $\balpha$ stage with self-attention, temporal information at $\bbeta$ stage with multiple kernels, and contextual information at $\bgamma$ stage with recurrent states, for sequential recommendation.}
  \Description{Model architecture}
  \label{fig:model_arch}
  \vspace{-3mm}
\end{figure}

% \textcolor{blue}{Renqin: I think we could add some discussion about the general picture of our model design first.} 
In this section, we will introduce from a high level about each part of our CTA model in a bottom-up manner, from the inputs, through the three stage pipeline: content-based attention, temporal kernels and contextualized mixture, denoted as $\balpha\rightarrow\bbeta\rightarrow\bgamma$ stages as illustrated in Figure \ref{fig:model_arch}, and finally into the output.

The raw input consists of the user's historical events of a window size $L$ in item and time pairs $ \{ (t_i, s_i) \}_{i=1}^{L}$, as well as the timestamp at the moment of recommendation $t_{L+1}$.  The sequence of input items is mapped into embedding space with the input item embeddings $ E_{\textup{input}} \in \mathbb{R}^{N\times d_{\textup{in}}} $: $\boldsymbol{X} = [s_1, \dots,  s_L]   \cdot E_{\textup{input}} \in \mathbb{R}^{L\times d_{\textup{in}}}$. 
We also transform the sequence of timestamps into the intervals between each action to current prediction time:  $\bT =  [t_{L+1} - t_1, \dots,  t_{L+1} - t_L] \in \mathbb{R}^{L\times 1}$.

%Now we determine the influence of each input through the Contextualized Temporal Attention Mechanism: 
Motivated by our earlier analysis, we design the three stage mechanism, namely $M^{\alpha}$, $M^{\beta}$ and $M^{\gamma}$, on top of the processed input $\boldsymbol{X}$ and $\bT$, to model dependencies among the historical interactions respectively on their content, temporal, and context information:% in the sequential recommendation setting:
\begin{equation*}   
    \balpha = M^{\alpha}( \bX ) \rightarrow \bbeta = M^{\beta}( \bT ) \rightarrow \bgamma = M^{\gamma}( \bX, \bbeta, \balpha ) 
\end{equation*}
In essence, $M^{\alpha}$ weighs the influence of each input purely on content $\bX$ and outputs a scalar score as importance of each events in sequence $\balpha \in \mathbb{R}^{L\times 1} $; $M^{\beta}$ transforms the temporal data $\bT$ through $K$ temporal kernels for the temporal weighing of each input $\bbeta\in \mathbb{R}^{L\times K}$; $M^{\gamma}$ extracts the context information from $\bX$, with which it mixes the factors $\balpha$ and $\bbeta$ from previous stages into the contextualized temporal importance score $\bgamma \in \mathbb{R}^{ L \times 1} $. We will later explain their individual architectures in details.  

In the end, our model computes the row sum of the input item sequence embedding $\bX$ weighted by $\bgamma$ (through the softmax layer, the weight $\bgamma $ sums up to 1). This weighted sum design is borrowed from the attention mechanism in a sense of taking expectation on a probability distribution, $\bgamma = \left[P(\hat{x}_{L+1} = x_i | \bX, \bT)\right]_{i=0}^{L}$.
%of next item representation is modeled as the weighted average of input items. 
The representation is then projected to the output embedding space $\mathbb{R}^{d_{\textup{out}} }$ with a feed-forward layer $F^{\text{out} }$  : %with activation function \cite{Nair:2010:RLU:3104322.3104425}
\begin{equation*}
\hat{x}_{L+1} = F^{\text{out} }(\bgamma^T \cdot \bX ) \in \mathbb{R}^{ d_{\textup{out}} \times 1} 
\end{equation*}
%where we let the value vector $\boldsymbol{V}_i$ be each of the input embedding $X_i$. 

We consider $\hat{x}_{L+1}$ as the predicted representation of recommended item. We define matrix $E_{\textup{output}} \in \mathbb{R}^{N\times d_{\textup{out}}}$, where its $i$th row vector is the item $i$'s representation in the output embedding space.  Then $\forall i\in \mathcal{I}$, the model can compute the similarity $r_i$ between item $i$ and the predicted representation $\hat{x}_{L+1}$ through inner-product (or any other similarity scoring function):
\begin{equation*} 
\hat{s}_{L+1} = (r_1, \dots, r_N) =   E_{\textup{output}} \cdot \hat{x}_{L+1} \in \mathbb{R}^{ N \times 1} 
\end{equation*}

For a given user, item similarity scores are then normalized by a softmax layer which yields a probability distribution over the item vocabulary. After training the model, the recommendation for a user at step $L+1$ is served by retrieving a list of items with the highest scores $r_v$ among all $ v\in \mathcal{V}$.
%The model can then compute the emission probability: 
%$$   P(\hat{s}_{l+1} \in \mathcal{I} ) = \operatorname{softmax}(H_{l} \cdot E_{\textup{output}}^T) \in R^{1\times N} $$
%$P(\hat{s}_{l+1} \in \mathcal{I} )$

% \textcolor{blue}{Renqin: I think we could add more discussions about why we design each component in this way.}
\subsection{Three Stage Weighing Pipeline}
\subsubsection{$\balpha$ stage, what is the action:} The goal of $\balpha$ stage is to obtain the content-based importance score $\balpha$ for the input sequence $\bX$. Following the promising results of prior self-attentive models, we adopt the self-attention mechanism to efficiently and effectively capture the content correlation with long-term dependence. In addition, the self-attention mechanism allows us to directly define the importance score over each input, in contrast to the recurrent network structure.

We use the encoder mode of self-attention mechanism to transform the input sequence embedding $\boldsymbol{X}$, through a stack of $d_l$ self-attentive encoder blocks with $d_h$ heads and $d_a$ hidden units, into representation $\boldsymbol{H}^{d_l}$, which is the hidden state of the sequence at the last layer. 
Due to the recursive nature of self-attention, we use the following example to explain the multi-head attention component in our solution. For example, in the $i$th attention head of the $j$th self attention block, from the input state $\boldsymbol{H}^j$, we compute one single head of the self-attended sequence representation as,
\begin{equation*}
   \boldsymbol{z}_i^j = \operatorname{Attention}( \boldsymbol{H}^j W_i^Q , \boldsymbol{H}^j W_i^K, \boldsymbol{H}^j W_i^V )
\end{equation*}
where  $ W_i^Q, W_i^K, W_i^V \in \mathbb{R}^{ d_a \times d_a / d_h } $ are the learnable parameters specific to $i$th head of $j$th attention block, used to project the same matrix $\boldsymbol{H}^j$ into the query $\bQ$, key $\bK$, and value $\bV$ representation as the input to   
 the Scaled Dot-Product \cite{vaswani2017attention}:
\begin{equation*} 
 \operatorname{Attention}(\bQ, \bK, \bV) = \operatorname{softmax}\left(\frac{ \boldsymbol{Q} \cdot \boldsymbol{K}^T}{\sqrt{d_a / d_h} } \right) \bV
\end{equation*}
Here the scaling factor $\sqrt{d_a / d_h}$ is introduced to produce a softer attention distribution for avoiding extremely small gradients. 

All the computed heads $\boldsymbol{z}_i^d$ in the $j$th attention block is stacked and projected as $\boldsymbol{Z}^j = [\boldsymbol{z}_1^d, \dots,\boldsymbol{z}_{d_h}^d] \cdot W^O $, where $W^O\in \mathbb{R}^{ d_a \times d_a } $. We can then employ the residue connection \cite{He2015DeepRL} to compute the output of this attention block as:
\begin{equation*} 
    \boldsymbol{H}^{j+1} = \operatorname{LN}\left( \boldsymbol{H}^{j} +  F^{j} ( \boldsymbol{Z}^j ) \right) 
\end{equation*}
where $F^{j}$ is a feed-forward layer specific to the $j$th attention block mapping from $ \mathbb{R}^{ d_a}$ to $\mathbb{R}^{  d_{\textup{in}} } $  and  $\operatorname{LN}$ is the Layer Normalization function \cite{ba2016layer}.
%defined as:
% \begin{equation}
%     \operatorname{LN}(x)=y \odot \frac{x-\mu}{\sqrt{\sigma^{2}+\epsilon}}+\beta
% \end{equation}

Note that for the initial attention block, we use $\bX$ to serve as the input $\bH^0$; and in the end, we obtain $\bH^{d_l}$ as the final output from self-attention blocks. In prior work \cite{kang2018self,sun2019bert4rec}, this $\bH^{d_l}$ is directly used for prediction. Our usage of self-attention structure is to determine a reliable content-based importance estimate of each input, hence we compute once again the Scale Dot-Product using the last layer hidden states $\boldsymbol{H}^{j+1}$ to project as the query and the last item  input embedding $x_L ^T$ to project as the key via $W_0^Q, W_0^K \in \mathbb{R}^{d_{\textup{in}} \times d_{\textup{in} } }$:
\begin{equation*} 
    \balpha = \operatorname{softmax}\left(  \frac{ (\boldsymbol{H}^{j+1} W_0^Q ) \cdot (x_L W_0^{K} )^T }{ \sqrt{d_{\textup{in}}} } \right)
\end{equation*}
Note that we can also view this operation as the general attention \cite{luong2015effective}, i.e., the bi-linear product of the last layer hidden states and the last input item embedding, where $ W_0^Q (W_0^{K} )^T$ is the learnable attention weight and $\sqrt{d_{\textup{in}}}$ serves as the softmax temperature \cite{hinton2015distilling}.

\subsubsection{$\bbeta$ stage, when did it happen:} The goal of $\bbeta$ stage is to determine the past events' influence based on their temporal gaps from the current moment of recommendation. The raw information of time intervals might not be as useful to indicate the actual \textit{temporal distance} of a historical event's influence (e.g., perceived by the user), unless we transform them with some appropriate kernel functions. 

Meanwhile, we incorporate the observation that each event can follow different dynamics in the variation of its temporal distance, given different contextual conditions. The item browsed casually should have its influence to user preference drop sharply for a near term, but it might still be an important indicator of user's general interest in the long term. In contrast, if the user is seriously examining the item, it is very likely the user would be interested to visit the same or similar ones in a short period of time.  Therefore, we create multiple temporal kernels to model the various temporal dynamics and leave it for the context environment to later decide contextualized temporal influence. This design allows more flexibility in weighting the influence of each event with different temporal distances.

In this paper, we handpicked a collection of $K$ kernel functions $\phi(\cdot): \mathbb{R}^L \rightarrow \mathbb{R}^L$ with different shapes including:
\begin{enumerate}
    \item exponential decay kernel, $\phi\left( \bT \right) = a e^{- \bT} + b$, assumes that the user's impression of an event fades exponentially but will never fade out. %the influence drops slower as time goes and converges to $b$
    \item logarithmic decay kernel, $\phi\left( \bT\right) =  -a \log (1+ \bT) + b$, assumes that the user's impression of an event fades slower as time goes and becomes infinitesimal eventually. Later we will introduce a softmax function that will transform negative infinity to $0$. %the influence drops slower as time goes, but still drops to $-\infty$.
    \item linear decay kernel, $\phi\left( \bT\right) =  -a  \bT + b$, assumes that the influence drops linearly and the later softmax operation will map the influence over some time limit to 0. 
    \item constant kernel,  $\phi\left( \bT\right) =  \mathbf{1}$, assumes that the influence stays static.
\end{enumerate}
where $a, b \in \mathbb{R}$ are the corresponding kernel parameters. Note that the above kernels are chosen only for their stability in gradient descent and well understood property in analysis. We have no assumption of which kernel is more suitable to reflect the actual temporal dynamics, and an ablation study of different combinations is presented in the following Section \ref{exp:kernel}. This mechanism should be compatible with other types of kernel function $\phi(\cdot)$ by design, and it is also possible to inject prior knowledge of the problem to set fixed parameter kernels.

Hence, given a collection of $K$ kernel functions,  $\left\{ \phi(\cdot)^1, \dots, \phi(\cdot)^K \right\}$, we transform $T$ into K sets of temporal importance scores: $\bbeta = \left[ \phi^1(T), \dots, \phi^K(T) \right]$, for next stage's use.

%We define the notion of temporal distance as the original time gap transformed by the kernel function following certain temporal dynamics trend.
%And we reweigh the self-attentive dependence by its temporal distance following the temporal dynamics kernel function $\phi^k(\cdot)$:
%where the softmax operation is applied across the entire input sequence of dimension size $L$ for each of the K different kernel functions.

\subsubsection{$\bgamma$ stage, how did it happen:} The goal of $\bgamma$ stage is to fuse the content and temporal influence based on the extracted context information. The core design follows the multiple sets of proposed temporal dynamics in the $\bbeta$ stage, in which it learns the probability distribution over each temporal dynamics given the context. 

First, we explain our design to capture context information. In our setting, we consider the contextual information as two parts: sensitivity and seriousness. Specifically, if one event seems to be closely related to its future actions, it means the user is likely impressed by this event and his or her ongoing preference should be sensitive to the influence of this action. In contrast, if the event appears to be different from its past actions, the user is possibly not serious about this action, since his or her preference does not support it. Such factors of sensitivity and seriousness can be valuable for the model to determine the temporal dynamics that each particular event should follow. Review the example in Figure \ref{fig:motiv_exp} again, the repetitive interactions with smartphones reflect high seriousness, while the sparse and possibly a noisy click on shoes suggests low sensitivity to its related products.  This observation also motivates our design to model context as its relation from past and to future events: we choose the Bidirectional RNN structure \cite{schuster1997bidirectional} to capture the surrounding event context from both directions. From the input sequence embedding $\bX$, we can compute the recurrent hidden state of every action as their context feature vector:
\begin{equation*}
    \boldsymbol{C} = \operatorname{Bi-RNN}(X) \oplus C_{\textup{attr}} \in \mathbb{R}^{L \times d_{r}} 
\end{equation*}
where $\oplus$ is the concatenation operation. Here, we also introduce some optional context features $C_{\textup{attr}}$ that can be the attributes of each event in the specific recommendation applications, representing the context when the event happened. For instance, we can infer the user's seriousness or sensitivity from the interaction types (e.g., purchase or view) or the media (e.g., mobile or desktop) associated with the action. In our experiments, we only use the hidden states of bidirectional RNN's output as the context features, and we leave the exploration of task specific context features as our future work.

Second, the model needs to learn the mapping from the context features of event $i$ to a weight vector of length $K$, where each entry $p_i(k|c_i)$ is the probability of this event follows $\phi^k(\cdot)$ as the temporal dynamics. 
We apply the feed-forward layer $F^{\gamma}$ to map them into the probability space $ \mathbb{R}^{K} $ and then normalize them into probabilities that sum up to one for each action with a softmax layer: 
\begin{equation*} 
    \boldsymbol{ P } ( \cdot | \boldsymbol{C}) = \operatorname{softmax}\left( F^{\gamma}( \boldsymbol{C} ) \right)
\end{equation*}

Finally, we use the probability distribution to mix the temporal influence scores from the $K$ different kernels for the contextualized temporal influence $\bbeta^c = \bbeta \cdot \boldsymbol{ P } ( \cdot | \boldsymbol{C} ) $, with which we use element-wise product to reweight the content-based importance score for the contextualized temporal attention score:
\begin{equation*}
\bgamma = \operatorname{softmax}\left( \balpha \bbeta^c   \right)
\end{equation*}
This design choice that uses product instead of addition to fuse the content and contextualized temporal influence score $\balpha$ and $\bbeta$ is based on the consideration of their influence on the gradients of $\theta^{\balpha}$. For example, the gradient on parameters in $\balpha$ stage is,
\begin{equation*}
\frac{ \partial ( \balpha \bbeta^c ) }{\partial \theta^{\balpha} } = \frac{ \partial \balpha  }{\partial \theta^{\balpha} }  \bbeta^c,
\quad
\frac{ \partial ( \balpha + \bbeta^c )  }{\partial \theta^{\balpha} } = \frac{ \partial \balpha  }{\partial \theta^{\balpha} }  
\end{equation*}
The error gradient in the addition form is independent of the function evaluation of $\bbeta^c$, while the product form has the gradients of $\balpha$ and $\bbeta^c$ depend on each other. Therefore, we choose the product form as a better fusion of the two scores.

\subsection{Parameter Learning}
\subsubsection{Loss Functions} In the previous section, we showed how the model makes recommendations by the highest similarity scores $\{r_v\}$ for all $v\in \mathcal{V}$. When training the model, we only use a subset of $\{r_v\}$. That is, since the size of the item space can be very large, we apply negative sampling \cite{mikolov2013distributed}, i.e., proportional to their popularity in the item corpus, sample a subset of items $N_{S} \subseteq  \mathcal{V}$, that excludes the target item $i$, i.e., $i\not\in N_{S}$.

We adopt negative log-likelihood (NLL) as the loss function for model estimation:
$$ L_{\textup{NLL}} = - \log \frac{e^{r_i}}{ \sum_{j\in N_{S}}^{} e^{r_j}}, $$
which maximizes the likelihood of target item.

We also consider two ranking-based metrics to directly optimize the quality of recommendation list. The first metric is the Bayesian Personalized Ranking (BPR) loss \cite{Rendle:2009:BBP:1795114.1795167} \footnote{ We use the sigmoid function $\sigma(x) = \frac{1}{1+e^{-x}}$.}:
$$ L_{\textup{BPR}} =  -\frac{1}{N_{S}} \sum_{j\in N_{S}}^{} \log \sigma\left(r_{i}-r_{j}\right), $$
which is designed to maximize the log likelihood of the target similarity score $r_{i}$ exceeding the other negative samples' score $r_{j}$. 

The second is the TOP1 Loss \cite{hidasi2018recurrent}:
$$ L_{\textup{TOP1}} = \frac{1}{N_{S}} \sum_{j\in N_{S}}^{} \sigma\left(r_{j}-r_{i}\right) + \sigma\left(r_{j}^2\right),  $$ 
which heuristically puts together one part that aims to push the target similarity score $r_{i}$ above the score $r_{j}$ of the negative samples, and the other part that lowers the score of negative samples towards zero, acting as a regularizer that additionally penalizes high scores on the negative examples. 

\subsubsection{Regularization} We introduce regularization through the dropout mechanism \cite{srivastava2014dropout} in the neural network. In our implementation, we have dropout layer after each feed-forward layer and the output layer of context bidirection RNN with a dropout rate of $0.2$. 
%% temporal kernel regularization
We leave as out future work to explore the effect of batch normalization as well as  regularization techniques of the parameters in temporal kernels.

\subsubsection{Hyperparameter Tuning} We initialize the model parameters through the Kaiming initialization proposed by \citet{he2015delving}. The temporal kernel parameters are initialized in proper range (e.g. uniform random in $[0,1]$) in order to prevent numerical instability during training.
We use the Relu function \cite{Nair:2010:RLU:3104322.3104425} by default as the activation function in the feed-forward layer.
%% Following the convention of learning rate scheduler in transformer model.
%% How parameters are initialized , temporal kernels

\section{Experiments}
In this section, we perform extensive experiment evaluations of our proposed sequential recommendation solution. We compared it with an extensive set of baselines, ranging from session-based models to temporal and sequential models, on two very large collections of online user behavior log data. We will start from the description of experiment setup and baselines, and then move onto the detailed experiment results and analysis.

\subsection{Dataset}
We use two public datasets known as XING and UserBehavior. The statistics of the datasets are summarized in Table \ref{table:Statistics}. The two datasets include user behaviors from two different application scenarios. 

\begin{table}[t]
\centering
\caption{Statistics of two evaluation datasets.}
\label{table:Statistics}
\vspace{-2mm}
\begin{tabular}{c|c|c} \hline
    Dataset & XING & UserBehavior  \\ \hline
    Users & 64,890 & 68,216\\
    Items & 20,662 & 96,438\\
    Actions & 1,438,096 & 4,769,051 \\
    Actions per user & 22.16$\pm$21.25 & 69.91$\pm$48.98 \\
    Actions per item & 69.60$\pm$112.63 & 49.45$\pm$65.31 \\
    Time span & 80 days & 9 days \\ \hline
\end{tabular}{}
\vspace{-2mm}
\end{table}

The XING dataset is extracted from the Recsys Challenge 2016 dataset \cite{Pacuk:2016:RCJ:2987538.2987544}, which contains a set of user actions on job postings from a professional social network site \footnote{https://www.xing.com/}. Each action is associated with the user ID, item ID, action timestamp and interaction type (click, bookmark, delete, etc.). Following the prior work \cite{quadrana2017personalizing,you2019hierarchical}, we removed interactions with type ``delete'' and did not consider the interaction types in the data. We removed items associated with less than 50 actions, and removed users with less than 10 or more than 1000 actions. We also removed the interactions of the same item and action type with less than 10 seconds dwell time.
%It contains interactions for 770k users over a 80-days period. %XING is a social network for business: people use XING to find a job and recruiters use XING to find the right candidate for a job. The user interactions come with timestamps and interaction types (click, bookmark, reply and delete). 

The UserBehavior dataset \cite{zhu2018learning} is provided by Alibaba and contains user interactions on commercial products from an e-commerce website\footnote{https://www.taobao.com/}. Each action is associated with the user ID, item ID, action timestamp and interaction type (click, favor, purchase, etc.). In order to have a computationally tractable deep learning model, we randomly sub-sampled 100,000 users' sequences from each dataset for our experiment. We removed items associated with less than 20 actions, and then removed users with less than 20 or more than 300 actions. We also removed the interactions with timestamp that is outside the 9 day range that dataset specifies.

\subsection{Experiment Setup}

\begin{table*}[t]
\centering
\caption{Performance comparison of different methods on sequential recommendation.}
\label{table:results}
\vspace{-2mm}
\begin{tabular}{|c|c|c|c|c|c|c|c|c|c|c|c} \hline
    Dataset & Metric & \textbf{CTA} & {Pop} & {S-Pop} & {Markov} & {GRU4Rec} & {HRNN}  & {LSHP} & {SASRec} & {M3R} \\ \hline
    XING & Recall@5 & \textbf{0.3217} & {0.0118} & {0.2059} & {0.2834} & {0.2690} & {0.2892}  & {0.2173} & {0.2530} & {0.2781} \\
        & MRR@5 & {0.1849} & {0.0062} & {0.1202} & {0.2319} & {0.2008} & {0.2392} & {0.1454}  & {0.2254} & \textbf{0.2469} \\ \hline
    UserBehavior & Recall@5 & \textbf{0.1611} & {0.0026} &  {0.1093} & {0.0846} & {0.0936} & {0.0940}  & {0.1201} & {0.1418} & {0.1077}\\
        & MRR@5 & \textbf{0.0925} & {0.0013}  & {0.0639} & {0.0534} & {0.0619} & {0.0610}  & {0.0792} & {0.0863} & {0.0689} \\ \hline
\end{tabular}{}
\vspace{-2mm}
\end{table*}

\subsubsection{Baseline methods}
We compare our proposed Contextualized Temporal Attention Mechanism with a variety of baseline methods\footnote{All implementations are open sourced at \url{https://github.com/Charleo85/seqrec}}. %but remain anonymous for double blind review} 
To ensure a fair comparison of deep learning model, we adjust the number of layers and hidden units such that all the models have similar number of trainable parameters. %We use no batch normalization for all model comparison experiments.

\textbf{Heuristics methods.} We include some simple heuristic methods, which show strong performance in prior sequential recommendation work \cite{quadrana2017personalizing, li2017neural}. 
\begin{itemize}[leftmargin=*]
    \item \textbf{Global Popularity (Pop).} Rank item by its popularity in the entire training set in a descending order.
    \item \textbf{Sequence Popularity (S-Pop).} Rank item by its popularity in the target user's action sequence in a descending order. The popularity of an item is updated sequentially as more actions of the target user are observed. 
    \item \textbf{First Order Markov Model (Markov).} This method makes the Markov assumption that each action depends only on the last action. It ranks item according to its probability given the item in last action, which is estimated from the training set. 
\end{itemize}

\textbf{Session-based Models.} We include several deep learning based models with session assumptions. We set the session cut-off threshold as 30 minutes by convention.
\begin{itemize}[leftmargin=*]
    \item \textbf{Session based Recurrent Neural network (GRU4Rec).} \citet{hidasi2015session} used the GRU, a variant of Recurrent Neural network, to model the user preference transition in each session. The session assumption is shown to be beneficial for a consistent transition pattern. %In the experiment, we use 500 hidden units.
    \item \textbf{Hierarchical Recurrent Neural network (HRNN).} \citet{quadrana2017personalizing} proposed a hierarchical structure that use one GRU to model the user preference transition in each session and another to model the transition across the sessions. %We set 300 hidden units for both the in- and cross-session GRU cells.
\end{itemize}

\textbf{Temporal Models.} Since our model additionally uses the temporal information to make the sequential recommendation, we include the following baselines that explicitly consider temporal factors and have been applied in sequential recommendation tasks. 
\begin{itemize}[leftmargin=*]
    % \item \textbf{Recurrent Marked Temporal Point Process (RMTPP).} \citet{du2016recurrent}
    \item \textbf{Long- and Short-term Hawkes Process (LSHP).} \citet{cai2018modeling} proposed a Long- and Short-term Hawkes Process that uses a uni-dimension Hawkes process to model transition patterns across sessions and a multi-dimension Hawkes process to model transition patterns within a session. 
    % \item \textbf{Recurrent Neural Network with Time Gates  (TRNN)} \citet{zhu2017next}
\end{itemize}

\textbf{Sequential Models.} Similar to our proposed CTA model, we also include several deep learning based models that directly learn the transition pattern in the entire user sequence. A fixed size window is selected for better performance and more memory-efficient implementation.
\begin{itemize}[leftmargin=*]
    \item \textbf{Self-attentive Sequential Recommendation  (SASRec).} \citet{kang2018self} applied the self-attention based model on sequential recommendation. It uses the last encoder's layer hidden state for the last input to predict the next item for user. We use 4 self-attention blocks and 2 attention heads with hidden size 500 and position embedding. We set the input window size to 8. 
    \item \textbf{Multi-temporal-range Mixture Model (M3R).} ~\citet{tang2019towards} proposed a mixture neural model to encode the users' actions from different temporal ranges. It uses the item co-occurrence as tiny-range encoder, RNN/CNN as short-range encoder and attention model as long-range encoder. Following the choice in its original paper, we use GRU with hidden size 500 as the short-range encoder. 
\end{itemize}

\subsubsection{Implementation Details} For our proposed model, we use self-attention blocks $d_l = 2$ and attention heads $d_h = 2$  with hidden size $d_a = 500$. We use the same representation for input and output item embeddings $E_{\textup{in}}=E_{\textup{output}}$, and a combination of 5 exponential decay kernels ($\psi^5$). We use a bidirection RNN with hidden size $d_r = 20$ in total of both directions to extract context features. We set the learning rate as $0.001$. We will present the experiments on different settings of our model in the following section.

% \begin{table*}[t]
% \centering
% \caption{Performance comparison of different methods on sequential recommendation.}
% \label{table:results}
% \begin{tabular}{c|c|c|c|c|c|c|c|c|c|c|c} \hline
%     Dataset & Metric & \textbf{CTA} & {Pop} & {S-Pop} & {Markov} & {GRU4Rec} & {HRNN}  & {LSHP} & {TRNN} & {SASRec} & {M3R} \\ \hline
%     XING & Recall@5 & \textbf{0.3217} & {0.0118} & {0.2059} & {0.2834} & {0.2690} & {0.2892}  & {0.2173} & {TRNN} & {0.2530} & {0.2781} \\
%         & MRR@5 & {0.1849} & {0.0062} & {0.1202} & {0.2319} & {0.2008} & {0.2392} & {0.1454} & {TRNN} & {0.2254} & \textbf{0.2469} \\ \hline
%     UserBehavior & Recall@5 & \textbf{0.1611} & {0.0026} &  {0.1093} & {0.0846} & {0.0936} & {0.0940}  & {0.1201} & {TRNN} & {0.1418} & {0.1077}\\
%         & MRR@5 & \textbf{0.0925} & {0.0013}  & {0.0639} & {0.0534} & {0.0619} & {0.0610}  & {0.0792} & {TRNN} & {0.0863} & {0.0689} \\ \hline
% \end{tabular}{}
% \end{table*}

\subsubsection{Experiment settings.} We split all the data by user, and select 80\% of the users to train the model, 10\% as the validation set and the remaining 10\% users to test the model. We also adopt the warm start recommendation setting, where the model is evaluated after observing at least 5 historical actions in each testing user.

All the deep learning based models are trained with Adam optimizer with momentum 0.1. We also search for a reasonablely good learning rate in the set $\{0.01, 0.001, 0.0001, 0.00001\}$ and report the one that yields the best results.  We set batch size to 100, and set the size of negative samples to 100. The model uses the TOP1 loss by default. The item embedding is trained along with the model, and we use the embedding size 500 for all deep learning models. The training is stopped when the validation error plateaus. For the self-attention based model, we follow the training convention \cite{zhu2018learning} by warming up the model in the first few epoches with small a learning rate.

\subsubsection{Evaluation metrics.} The model predicts the user action at the time of the next observed action. The result is evaluated by ranking the ground-truth action against a pool of candidate actions. For both datasets, the candidate pool is the set of all items in the dataset, though only a subset of negative items is sampled for model optimization. 

We rank the candidates by their predicted probabilities and compute the following evaluation metrics:
\begin{itemize}[leftmargin=*]
    \item \textbf{Recall@K.} It reports the percentage of times that the ground-truth relevant item and ranked within the top K list of retrieved items. %Higher Recall@K implies more accurate prediction.
    \item \textbf{MRR@K.}  The mean reciprocal rank is used to evaluate the prediction quality from the predicted ranking of relevant items. It is defined as the average reciprocal rank for ground-truth relevant items among the top K list of retrieved items. If the rank is larger than K, the reciprocal rank is 0. %Higher MRR@K implies more accurate prediction.
\end{itemize}

\subsection{Experimental results}
\subsubsection{Overall Performance.}
We summarize the performance of the proposed model against all baseline models on both dataset in Table \ref{table:results}. The best solution is highlighted in bold face. %As we can observe from the table, 

Similar to the results reported in prior work \cite{quadrana2017personalizing, kang2018self}, heuristics methods do show strong baseline performance on our sequential recommendation tasks. And based on their strong performance, we can conclude that the XING dataset features first order transition, while the UserBehavior dataset features sequence popularity. This results are not surprising because it is common for a user to visit the same item several times back and forth in online shopping scenarios, and to visit the next job posting closely related to the recent history. This also results in  different strengths of each model on both datasets, which we will analyze in the next two sections. 

Notably, on both datasets, our proposed model CTA outperforms all baselines in Recall@5 by a large margin (11.24\% on XING dataset, 14.18\% on UserBehavior dataset). The model's MRR@5 performance is strong on UserBehavior dataset, but weak on XING dataset. %Why is our model bad at MRR?
This suggests that our model fails to learn a good ranking for the first order transition pattern, since it uses a weighted sum of input sequence for prediction. Nevertheless, such weighted sum design is powerful to capture the sequential popularity pattern. It also shows that our model outperforms the self-attentive baselines, which suggests our design of the contextual temporal influence reweighing, i.e., $\balpha\bbeta^c$, improves sequential order modeling in recommendation applications, compared to the positional embedding borrowed from natural language modeling.

\subsubsection{Results on XING dataset.}
%Why is the RNN performance worse than first order markov model on XING dataset, worse than sequence popularity baseline on Userbehavior dataset.
The RNN-based methods outperformed both temporal models and attention-based models. This again confirms that the recurrent model is good at capturing the first order transition pattern or the near term information. We also observe that the hierarchical RNN structure outperforms the first order baseline, while the session-based RNN performs not as well as this strong heuristic baseline. This demonstrates the advantage of hierarchical structure and reinforces our motivation to segment user history for modeling users' sequential behaviors.

\subsubsection{Results on UserBehavior dataset.}
On the contrary to the observations on XING dataset, the temporal models and attention-based models outperformed RNN-based methods. This means the recurrent structure is weak at learning the sequential popularity pattern, while the attention-based approach is able to effectively capture such long-term dependence. Such conflicting nature of existing baselines is exactly one of the concerns this work attempts to address. This again validates our design to evaluate and capture the long- and short-term dependence through the proposed three stage weighing pipeline.

\subsection{Performance analysis}

\subsubsection{Ablation Study}
\begin{table}[t]
\centering
\caption{Ablation analysis on two datasets under metrics of Recall@5 (left) and MRR@5 (right). The best performance is highlighted in bold face. $\downarrow$ and $\uparrow$ denote a drop/increase of performance for more than 5\%. $\psi$, $\rho$, $\pi$, $\omega$ respectively denote the exponential, logarithmic, linear and constant temporal kernels. The superscript on the kernel function denotes the number of such kernel used in the model. }
\label{table:Ablation}
\begin{tabular}{l|lr|lr|l} \hline
\multicolumn{2}{c}{\multirow{2}{*}{Architecture}}  & \multicolumn{4}{c}{Dataset} \\ \cline{3-6}
    \multicolumn{2}{c}{} & \multicolumn{2}{c}{XING} & \multicolumn{2}{c}{UserBehavior} \\ \hline
    % \multicolumn{2}{c}{} & Recall@5 & MRR@5  & Recall@5 & MRR@5 \\ \hline
    \multicolumn{2}{c}{Base}  & 0.3216 & 0.1847 & 0.1611 & 0.0925 \\ \hline
    Window &    4 & 0.3115$\downarrow $ & 0.2167$\uparrow$  & 0.1488$\downarrow$ & 0.0899 \\ 
    size (L) & 16 & 0.3049$\downarrow $ & 0.1733$\downarrow $  & 0.1433$\downarrow$ & 0.0914 \\ 
             & 32 & 0.3052$\downarrow $ & 0.1735$\downarrow $  & 0.1401$\downarrow$ & 0.0950 \\ \hline
    Attention       & 1  & 0.3220 & 0.1851 & 0.1631 & 0.0926 \\ 
    blocks ($d_l$)  & 4  & 0.3217 & 0.1849 & 0.1631 & 0.0924 \\ \hline
    Attention      & 1  & 0.3225 & 0.1860 & 0.1622 & 0.0919 \\ 
    heads ($d_h$)  & 4  & 0.3225 & 0.1860 & 0.1646 & 0.0940 \\ \hline
    \multicolumn{2}{l}{$\neg$ Sharing embedding} & 0.1263$\downarrow $  & 0.0791$\downarrow $ & 0.1042$\downarrow $ & 0.0192$\downarrow $ \\ \hline
    Embedding        & 300 & 0.3147 & 0.1831 & 0.1622 & 0.0920 \\ 
    size ($d_{\textup{in}}$) & 1000 & 0.3207 & 0.1857 & 0.1628 & 0.0921 \\ \hline
    Loss    &  NNL   & 0.3130 & 0.1806 & 0.1571 & 0.0895 \\ 
    function &  BPR & 0.3163 & 0.1804 & 0.1598 & 0.0913 \\ \hline
    \multicolumn{2}{l}{Flat attention }  & 0.3215  & 0.1869   & 0.1588 & 0.0907  \\ \hline
    \multicolumn{2}{l}{Global context $\boldsymbol{ P } ( \cdot )$ }    & 0.3207  & 0.1839 & 0.1603 & 0.0912 \\ 
    \multicolumn{2}{l}{Local context $\boldsymbol{ P } ( \cdot | x )$ }    & 0.3210 & 0.1841 & 0.1591 & 0.0912 \\ \hline
    Kernel & $\omega^1$  & 0.3191 & 0.1827  & 0.1604 & 0.0910 \\ 
    types & $\psi^1$    & 0.3122 & 0.2141$\uparrow$   & 0.1591 & 0.0907 \\ 
    &  $\psi^{10}$ & 0.3207 & 0.1844 & 0.1627 & 0.0925\\ 
    & $\pi^1$     & 0.2917$\downarrow $ & \textbf{0.2323}$\uparrow$  & 0.1562  & 0.0976 \\ 
    & $\pi^5$     & 0.3025$\downarrow $ & 0.2209$\uparrow$  &  {0.1670}  & \textbf{0.1010}$\uparrow$ \\
    & $\pi^{10}$    & 0.3214 & 0.2183$\uparrow$ & \textbf{0.1673} & {0.0997}$\uparrow$ \\ 
     & $\rho^{5}$  & 0.3111 & 0.2196$\uparrow$ & 0.1618 & 0.0931 \\ 
     & $\rho^{10}$  & 0.3230 & 0.1869 & 0.1635 & 0.0932 \\ 
    & $\psi^{5},\rho^{5}$   & 0.3241 & 0.1888 & 0.1635 & 0.0932 \\ 
    & $\psi^{5},\pi^{5}$  & \textbf{0.3273} & 0.2146$\uparrow$ & \textbf{0.1673} & {0.0997}$\uparrow$ \\ 
    & $\psi^{5},\rho^{5},\pi^{5}$ & 0.3254 & 0.1971$\uparrow$ & 0.1664 & 0.0983$\uparrow$ \\ 
    \hline
\end{tabular}{}
\vspace{-3mm}
\end{table}

we perform ablation experiments over a number of key components of our model in order to better understand their impacts. Table \ref{table:Ablation} shows the results of our model's default setting and its variants on both datasets, and we analyze their effect respectively:

\textbf{Window size.} We found that the window size of $8$ appears to be the best setting among other choices of input window size among $\{4, 16, 32\}$ on both datasets. The exceptions are a smaller window size on XING and a larger window size on UserBehavior can slightly improve MRR@5, even though Recall@5 still drops. The reason might be suggested by the previous observation that the first order transition pattern dominates XING dataset so that it favors a smaller input window, while the sequence popularity pattern is strong in UserBehavior dataset such that it favors a larger input window size. 

\textbf{Loss functions.} The choice of loss function also affects our model's performance. The ranking based loss function, BPR and TOP1, is consistently better than the NLL loss, which only maximizes the likelihood of target items. The TOP1 loss function with an extra regularizer on the absolute score of negative samples can effectively improve the model performance and reduce the over-fitting observed in the other two loss functions.

\textbf{Self-Attention settings.} We compare the model performance on different $d_l$ and $d_h$ settings. The performance difference is minimal on XING, but relatively obvious on UserBehavior. This indicates the content-based importance score is more important in capturing the sequential popularity than first order transition pattern.
    
\textbf{Item embedding.} We test the model with separate embedding space for input and output sequence representations; and the model performance drops by a large margin. Prior work, e.g., \cite{kang2018self}, in sequential recommendation found similar observations. Even though the separate embedding space is a popular choice in neural language models \cite{press-wolf-2017-using}, but the item corpus appears to be more sparse to afford two distinct embedding spaces. The dimensionality of the embedding space, $d_{\textup{in}}=d_{\textup{out}}$ slightly affects the model performance on both datasets, and it at the same time increases Recall@5 and decreases MRR@5 score, and vice versa. A trade-off on ranking and coverage exists between larger and smaller embedding spaces.

\subsubsection{Discussion on Model Architecture}
To further analyze the strength and weakness of our model design, we conduct experiments specifically to answer the following questions:

\textbf{Does the model capture the content influence $\balpha$?} 
    %We compare the performance of transformer with average embedding that has flat attention to address the hardness in capturing content correlation
    To understand if our model is able to learn a meaningful $\alpha$, we replace the $M^{\alpha}$ component with a flat attention module, such that it always outputs $\balpha =  \textbf{1}$. And we list this model's performance in Table \ref{table:Ablation} as `Flat Attention'. 
    
    The performance stays almost the same on XING, but drops slightly on the UserBehavior dataset. It shows that the content-based importance score is less important for the sequential recommendation tasks when the first order transition pattern dominates, but is beneficial for the sequential popularity based patterns. It also suggests that contextualized temporal importance along is already a strong indicator of historical actions about current user preference.  

\textbf{Does the model extract the context information in $\bgamma$ stage?} \label{exp:kernel}
    As the effect of temporal influence depends on our context modeling component, we design the following experiments on the context component to understand the two follow-up questions.
    
    First, \emph{whether the local context information of each event is captured}. We replace the local conditional probability vector $\boldsymbol{ P } ( \cdot | \boldsymbol{C})$ with a global probability vector $\boldsymbol{ P } ( \cdot )$, i.e., a single weight vector learnt on all contexts. This model's performance is listed in the table as `Global Context'. We can observe a consistent drop in performance in both datasets. 
    
    Second, \emph{whether the local context is conditioned on its nearby events}. We replace the local conditional probability vector $\boldsymbol{ P } ( \cdot | \boldsymbol{C})$ with a local probability vector conditioned only on the event itself, $\boldsymbol{ P } ( \cdot | x )$. More specifically, instead of using the bidirectional RNN component, the model now uses a feed-forward layer to map each $x_i$ to the probability space $ \mathbb{R}^{K} $. This model's performance is listed in the table as `Local Context'.  We again observe a consistent drop in performance, though it is slightly better than the global context setting. 
    
As a conclusion, our model is able to extract the contextual information and its mapping into probability for different temporal influences on both datasets.

\textbf{Does the model capture temporal influence $\bbeta$?}
    We conduct multiple experiments on the number of temporal kernels and the combined effect of different kernel types.%, listed in table \ref{table:Ablation}. 
    
    Firstly, we want to understand the advantages and limitations of each kernel type. We look at the model performance carried out with a single constant temporal kernel $\omega^1$. Its performance on MRR@5 is the worst among all the other kernel settings on both datasets. At the same time, we compare the settings of 10  exponential $\psi^{10}$, logarithmic $\rho^{10}$ and linear $\pi^{10} $ kernels each. The 10 linear kernels setting is overall the best  on both datasets, especially in improving the ranking-based metrics. It shows that it is beneficial to model the temporal influence with the actual time intervals transformed by appropriate kernel functions.  

    Secondly, we compare the model performance on different number of temporal kernels. The results suggested that the model performance always improves from using a single kernel to multiple kernels. This directly supports our multi-kernel design. Specifically, among the exponential kernels $\{\psi^1, \psi^5, \psi^{10}\}$, $\psi^5$ performs the best on XING, yet not as good as $\psi^{10}$  on UserBehavior. On linear kernels $\{\pi^1,  \pi^{5}, \pi^{10}\}$, as the kernel number increases, Recall@5 improves, but MRR@5 drops on XING. Similarly on UserBehavior, $\pi^{5}$ achieves the best ranking performance, but the $\pi^{10}$ induces a better coverage. Hence, the model with more kernels does not necessarily perform better, and as a conclusion, we need to carefully tune the number of kernels for better performance on different tasks. 
    
    Thirdly, we study the combinatorial effect of different kernel types: $(\psi^{5},\rho^{5})$,  $(\psi^{5},\pi^{5})$ and $(\psi^{5},\rho^{5},\pi^{5})$. We can observe that all types of kernel combinations we experimented improve the performance on both datasets, compared to the base setting $\psi^{5}$. This suggests the diversity of kernel types is beneficial to capture a better contextualized temporal influence. However, it also shows on both datasets that if mixing exponential $\psi^{5}$ with either linear $\pi^{5}$ or logarithmic $\rho^{5}$ kernel can improve the model performance, mixing all three of them together would only worsen the performance. We hypothesize that certain interference exists among the kernel types so that their performance improvement cannot simply add on each other. And we leave the exploration of finding the best combination of kernels as our future work.

Overall, we believe that the temporal influence can be captured by current model design, and there are opportunities left to improve the effectiveness and consistency of the current kernel based design.

% \textcolor{blue}{Renqin: I think we may try to select some example sequences to look into what of contextual information our model can capture if possible. This is generally tricky to select. But I think we could give it a try. At least show what kind of attention we can learn.}
\begin{figure*}[t]
  \centering
  \includegraphics[width=0.33\linewidth]{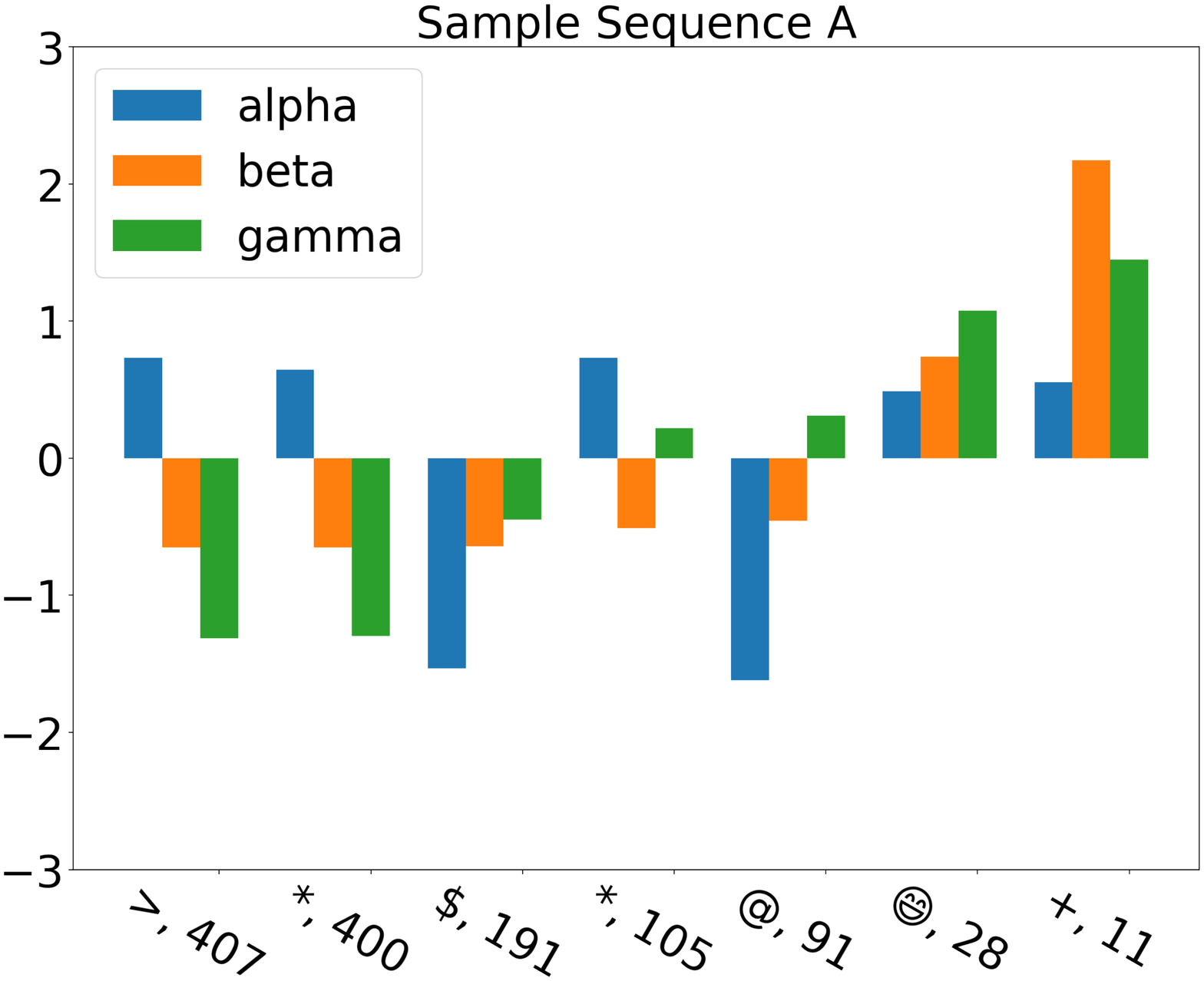}
  \includegraphics[width=0.33\linewidth]{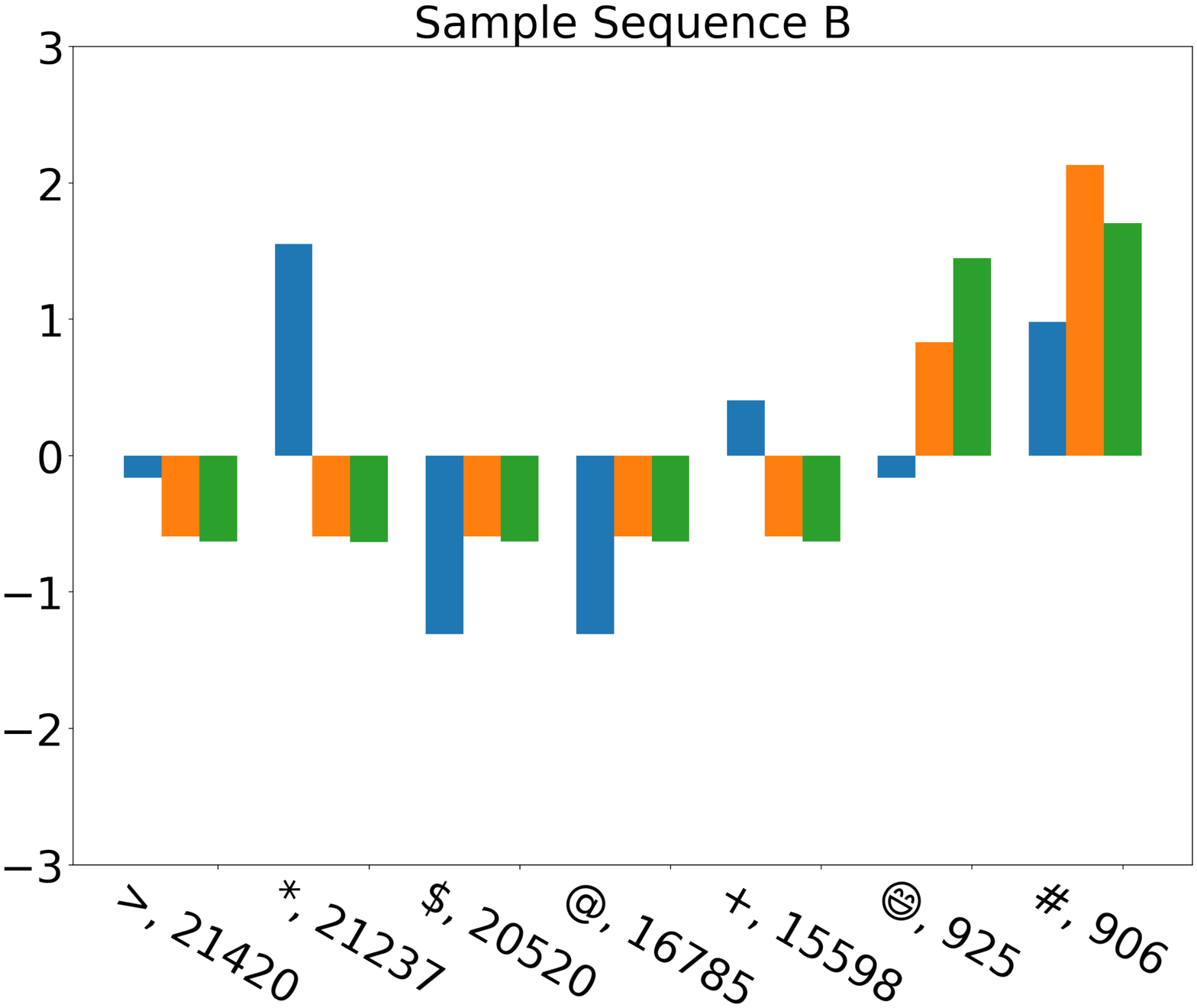}
  \includegraphics[width=0.33\linewidth]{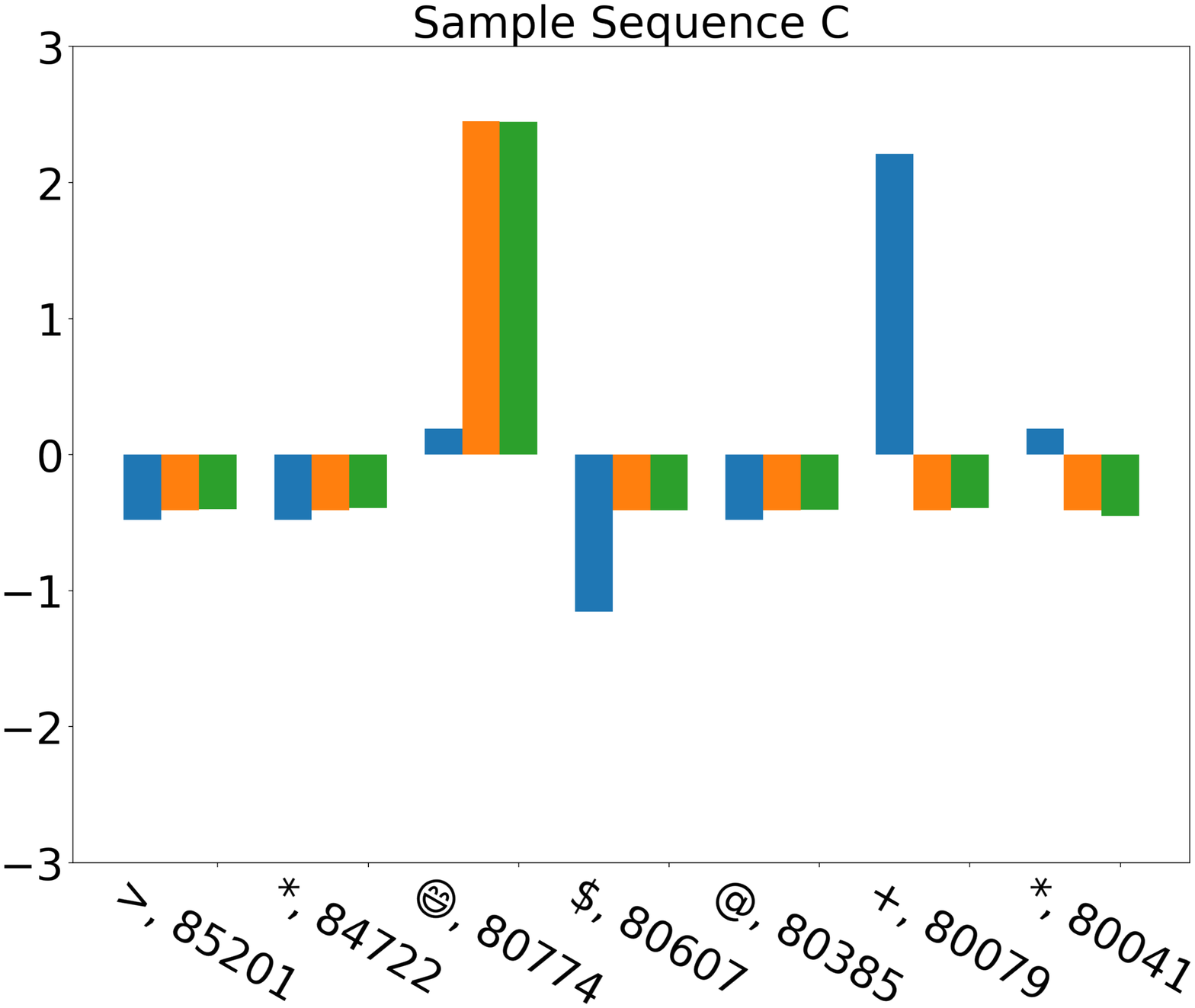}
  \caption{Attention visualization. The blue (left) bar is the content-based importance score $\balpha$, the orange (middle) bar is the contextualized temporal influence score $\bbeta^c$, the green (right) bar is the combined importance score $\bgamma$. The figures contains three different sequences selected from the test set of the UserBehavior dataset. }
  \Description{Model architecture}
  \label{fig:att_vis}
%   \vspace{-3mm}
\end{figure*}

\subsubsection{Attention Visualization}
To examine the model's behavior, we visualize how the importance score shifts in some actual examples in Figure \ref{fig:att_vis}. The x-axis is a series of actions with their associated items \footnote{for privacy concerns, these datasets do not provide the actual item content; and we represent the items in the figure with symbols.} and the time interval from action time to current prediction time, $(s_i, t_{L+1}-t_i)$. From left to right, it follows a chronological order from distant to recent history. We select the example such that the ground-truth next item is among the historical actions for the sake of simplicity, and we use smile face symbol \Laughey to denote if the item of such historical action is the same as the target item. Each action on the x-axis is associated with three bars. Their values on the y-axis is presented as the computed score $\balpha$, $\bbeta^c$ and $\bgamma$ respectively of each event in the model after normalization (by z-score). The model setting uses the temporal kernel combination $(\psi^{5},\pi^{5})$ for its best performance.

\textbf{Orange bars.} The contextualized temporal influence score $\bbeta^c$, in both sequence A and B, follows the chronological order, i.e., the score increases as time interval shortens. In addition, such variation is not linear over time: the most recent one or two actions tend to have higher scores, while the distant actions tend to have similar lower scores. The sequence C, as all actions happened long time ago from current prediction, the context factor is deciding the height of orange bar. And the model is able to extract the context condition and assign high temporal importance to this event, which is indeed the target item. These observations all suggest that the contextualized temporal influence is captured in a non-trivial way that helps our model to better determine the relative event importance.  

\textbf{Blue bars.} For the content-based importance score $\balpha$, it shows different distribution on each of the sequences. This is expected as we want to model the importance on the event correlation that is independent of the sequence order. Only in the third example that the target, i.e., the most relevant historical action, is ranked above average according to the content-based importance score. This again shows the important role of the temporal order to improve the ranking quality for sequential recommendation.

\textbf{Green bars.} The combined score $\gamma$ largely follows the relative importance ranking in orange bar. In other words, the contextualized temporal order is the dominating factor to determine relative importance of each input in our selected examples. This corresponds to the previous observation that the model performance would only slightly drop if the self-attention component outputs flat scores. This supports our motivation to model the contextualized temporal order in sequential recommendation tasks. 

Although these are only three example interaction sequences from more than $6,000$ users, we can now at least have a more intuitive understanding of the reweighing behavior of our model design -- the core part that helps boost the recommendation performance over the existing baselines. However, there are also many cases where the importance scores are still hard to interpret, especially if there is no obvious correspondence between target item and the historical actions. We need to develop better techniques to visualize and analyze the importance score for interpretable neural recommender system as follow-up research.

\section{Conclusion}
This work identifies and addresses the critical problem in sequential recommendation, \emph{Déjà vu}, that is the user interest based on the historical events varies over time and under different context. Our empirical evaluations show that the proposed model, CTA, has the following advantages:
\begin{itemize}[leftmargin=*]
    \item \textbf{Efficacy \& Efficiency.} Compared with the baseline work, CTA effectively improves the recommendation quality by modeling the contextualized temporal information.  It also inherits the advantage of self-attention mechanism for its reduced parameters and computational efficiency, as the model can also be deployed in parallel.
    \item \textbf{Interpretability.} Our model, featuring the three stage weighing mechanism, shows promising traits of interpretability. From the elementary analysis demonstrated in our experiments, we can have a reasonable understanding on why an item is recommended, e.g., for its correlation with some historical actions and how much on temporal influence or under context condition.
    \item \textbf{Customizability.} The model design is flexible in many parts. In the $\balpha$ stage, the model can extract the content-based importance by all means, such as the sequence popularity heuristics -- customizable for recommendation applications with different sequential patterns. In the $\bbeta$ stage, as we mentioned earlier, we can adapt different choices of temporal kernels to encode prior knowledge of the recommendation task. The $\bgamma$ stage is designed to incorporate extra context information from the dataset, and one can also use more sophisticated neural structures to capture the local context given the surrounding events.
\end{itemize}

Nevertheless, our understandings are still limited in the temporal kernels including what choices are likely to be optimal for certain tasks, and how we can regularize the kernel for more consistent performance. Our current solution ignores an important factor in recommendation: the user, as we assumed everything about the user has been recorded in the historical actions preceding the recommendation. As our future work, we plan to explicitly model user in our solution, and incorporate the relation among users, e.g., collaborative learning, to further exploit the information available for sequential recommendation.
%We will also work to exploit the interpretability of our model through the importance scores and explore other interpretable machine learning techniques to integrate in our model design.

\section{Acknowledgements}
We thank all the anonymous reviewers for their helpful comments. This work was partially supported by the National Science Foundation Grant IIS-1553568.

\bibliographystyle{ACM-Reference-Format}
\bibliography{mybib}

\end{document}